\documentclass[11pt,twoside]{article}

\usepackage{asp2006}
\usepackage{epsf}

\begin{document}
\title{Dark Lessons from the Sloan Digital Sky Survey}   
\author{Robert Nichol}   
\affil{Institute of Cosmology and Gravitation (ICG), Univ. of Portsmouth, Portsmouth, PO1 2EG}

\begin{abstract} 
The true nature of dark energy remains
unclear: It is either a strange fluid in the Universe, with a negative
effective pressure, or a breakdown in General Relativity on large
scales. This question can only be answered through a suite of
different observations as a function of redshift. In this paper, I
will briefly review our attempts to achieve this goal using data from the Sloan
Digital Sky Survey (SDSS). In particular, I will present new measurements of the Baryon Acoustic Oscillations (BAO) from the SDSS DR5 galaxy redshift survey as well as outline the on--going SDSSII Supernova survey, which has already detected (in 2005--06) over 300 SNe Ia's over the redshift range $0.05<z<0.4$. I will also discuss the latest measurements of the Integrated Sachs--Wolfe (ISW) effect that now probe the density of dark energy at $z\sim1.5$. All these measurements are still consistent with a $\Lambda$-dominated universe. 

\end{abstract}

\section{Introduction}

The most striking discovery in astrophysics over the last ten years is
that the energy density of the Universe is dominated by a mysterious
quantity called ``dark energy'' (Spergel et al. 2006). This dark
energy is responsible for a late-time acceleration of the Hubble
expansion of the Universe and its exact nature remains unclear. In
its simpliest form, dark energy could be Einstein's Cosmological
Constant ($\Lambda$), yet its observed value is substantially smaller
than expected for the vaccum energy density. Dark energy could be another scalar field that evolves with cosmic time, e.g.,
Quintessence. Alternatively, the late-time acceleration of the Universe could be the result of our lack of understanding of gravity
on large--scales, and many authors have recently proposed modified
gravity models to account for these cosmological observations
(e.g. Fairbairn \& Goobar 2005; Maartens \& Majerotto 2006).

The next decade will be dominated by new efforts to measure dark
energy to greater precision and therefore, determine its true nature. In particular, we can attempt to answer two fundamental questions about dark energy:

\begin{enumerate}

\item {\it Is dark energy just the Cosmological Constant?} This
question will be addressed through accurate measurements of the
equation of state of dark energy ($p = w \rho \, c^2$) as a function of cosmological
time. A cosmological constant is given by $w=-1$, while quintessence models usually predict values in the range of $-1<w\le 0$.

\item {\it Is dark energy a modification of gravity?} This question will likely be addressed through probing the Universe using different methods and tracers of the dark energy? In particular, we may expect differences in the rate of growth of cosmic structures in the Universe (see Ishak, Upadhye \& Spergel 2006; Linder 2006; Huterer \& Linder 2006)

\end{enumerate}

These issues have been extensively discussed in a series of recent dark energy reviews by both national and international organizations. In particular, I highlight below three outstanding reviews of the dark energy physics and summarise their key recommendations.

\begin{itemize}

\item{{\it The Dark Energy Review} by Trotta \& Bower (astro--ph/0607066) which was commissioned by the Science Committee of PPARC. This report clearly favors weak lensing and baryon acoustic oscillations (BAO) experiments to understand dark energy. This is primarily due to their statistical accuracy and robustness to systematic uncertainties.}

\item{{\it The Dark Energy Task Force (DETF)} by Albrecht et al. (astro-ph/0609591) in the US. This is the most comprehensive of the reports providing a quantitative ``figure of merit" for the various dark energy experiments and techniques. Their major recommendations include the use of multiple techniques to study dark energy, with at least one of these techniques being a probe of the  growth of structure in the Universe. They recommend immediate funding for projects that improved our understanding of systematics in the dark energy measurements, as these are now the dominant source of uncertainty.}

\item{{\it The ESA-ESO Working Group on Fundamental Physics} report by Peacock et al. The report focuses on european projects that could make significant  progress in understanding dark energy. The report recommends the undertaking of a space-borne imaging survey over a major fraction of the sky, complemented by photometric redshifts from new optical and infrared ground--based surveys. The report also highlights the importance of new spectroscopic surveys of $>10^5$ galaxies to calibrate the photometric redshifts. The report also notes the importance of improved local samples of supernovae to fully exploit them as cosmological probes. }

   \begin{figure}[tp]
\plotone{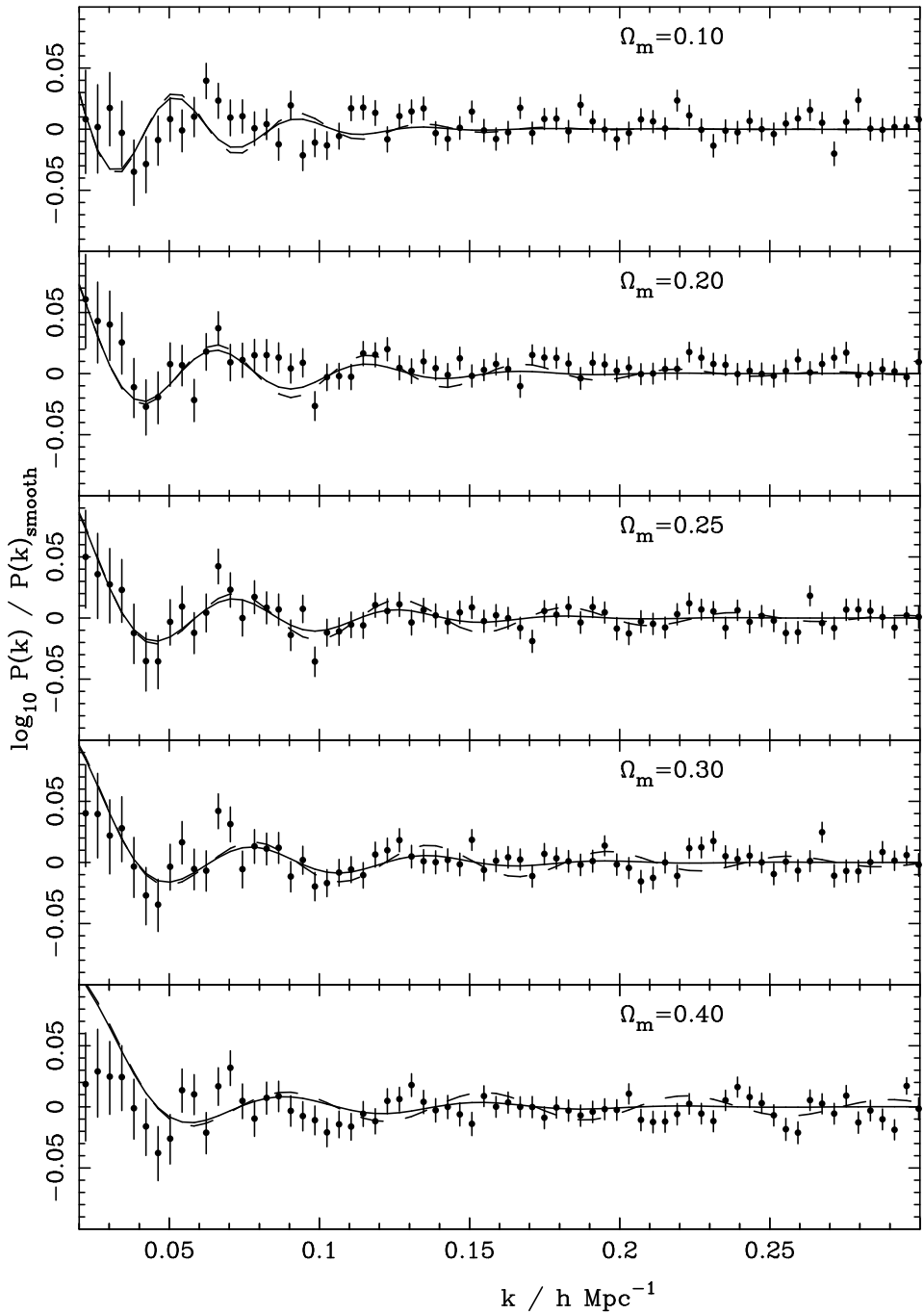}
\caption{Figure taken from Percival et al. (2006a) showing the ratio of the power spectra calculated from the SDSS to a smooth cubic spline fit used to remove the overall shape. The data are plotted as solid circles (with 1 sigma errors) using five flat $\Lambda$
    cosmological models to convert from redshift to comoving distance,
    with matter densities given in each panel. For comparison, we also plot the BAO predicted by a $\Lambda$CDM model with the same
    matter density, $h=0.73$, and a 17\% baryon fraction (solid
    lines). As can be seen, the
    observed oscillations approximately match those predicted by this
    model for $0.2\le\Omega_M\le0.3$.)
            \label{fig:percival}
           }
    \end{figure}

\end{itemize}

In general, all these reports highlight the 
need for new massive surveys of the Universe using dedicated
facilities. This is simply due to the fact that dark energy is a small
observational signal and thus requires big surveys to beat the
statistical noise (cosmic variance and shot--noise) and new
experiments to control the systematic uncertainties. Today, weak lensing (WL) and cluster
surveys are uncompetitive (compared to the existing BAO and SNe
surveys) but all the reports highlight that this will change in the coming
years, with WL surveys having the most promise. 

This situation is clearly detailed in the DETF report which
tabulates the fractional decrease in the error ellipse on the equation of state of dark energy as
a function of experiment type (BAO, SNe, clusters, WL). In particular,
they compare the error ellipses from the expected
pre--2010 experiments (called Stage I \& II experiments in their report, and
represent surveys already funded and underway), to the post-2010
experiments (Stage III \& IV), that are now being considered for funding around
the world. They also provide optimistic and pessimistic estimates for
the fractional gain in the errors to encompass our present understanding of the systematic uncertainties in each of these techniques.

Generally, the BAO and SNe methods have close optimistic and pessimistic
estimates indicating that these are now mature techniques with understood systematics. In constrast, the optimistic and pessimistic errors for WL
and clusters are widely discrepant reflecting the uncertainty with
these methods. However, the potential for large fractional increases
in our knowledge (in the optimistic case) is greatest for these two
techniques. In other words, WL and clusters offer the {\it ``high risk,
high gain''} options, while BAO and SNe are the {\it ``safe''} options
(although these techniques  will still deliver $>100$\% improvement in the errors on the
dark energy parameters after 2010).

In summary, which of these techniques should the astronomical community pursue? The
answer is the same advice one would receive when considering ones retirement funds, i.e., diversify. Some of our assumptions about the
systematic errors will likely be wrong, so we need ``safe" options to
spread this risk. Meanwhile, the riskier options provide orthogonal
information and therefore, will lead to a greater understanding beyond
the simple sum of the parts. 

  \begin{figure}[tp]
\plotone{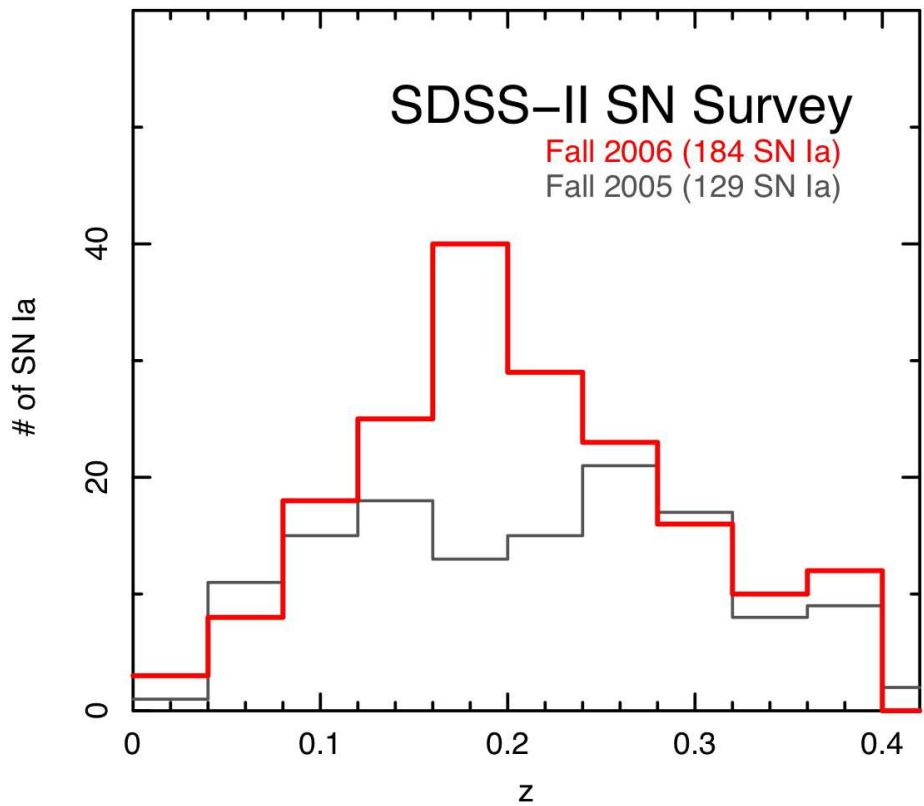}
\caption{The redshift histogram of spectroscopically--confirmed supernovae Type Ia's detected in 2005 (thin grey line) and 2006 (thicker red line) as part of the SDSSII Supernova Survey.
            \label{SNe}
           }
    \end{figure}

\section{Baryon Acoustic Oscillations}
 
The baryon acoustic oscillations (BAO) have received considerable
attention over the last 5 years and have emerged as a key technique
for measuring $w(z)$ as outlined in the reports above. The BAO are
caused by sound waves propogating through the primodial plasma in the
early Universe. At recombination, these sound waves are frozen into
the distribution of matter as a prefered scale given by the
$\simeq0.57ct$, where $c$ is the speed of light and $t$ is the age of
the Universe since the Big Bang. Therefore, the BAO represent a
standard ruler in the Universe, which is left imprinted in the
distribution of matter. See Eisenstein \& Hu (1998) for a
comprehensive review of the physics of the BAO, or Bassett, Nichol \& Eisenstein (2005) for a popular review of the BAO.

The BAO standard ruler has already been measured at the surface of last scattering
as the Doppler peaks in the CMB power spectrum. This
provides an accurate estimation for the distance to this surface (Spergel
et al. 2006). Clearly, if one can detect and measure the BAO at other
redshifts, then one can jointly constrain the geometry of the Universe
and its content as a function of redshift.

In the last 5 years, there have been several measurements of the BAO in the
distribution of galaxies in the late Universe. In 2001, Miller, Nichol \& Batuski (2001) and Percival et al. (2001) presented first evidence for the
BAO in the Abell cluster catalogue and 2dFGRS respectively. For example, Miller et
al. (2001) obtained constraints on the cosmological parameters that are fully consistent with the present--day best--fit
cosmology (Spergel et al. 2006; Percival et al. 2006b; Tegmark et
al. 2006). In 2005, both the SDSS and 2dFGRS provided convincing
evidence for the BAO in the distribution of local galaxies (see
Eisenstein et al. 2005; Cole et al. 2005). In 2006, Percival et al. (2006a)
presented a detailed analysis of the SDSS DR5 galaxy redshift survey
and provides a $3\sigma$ detection of the BAO signal independent of the shape
of the power spectrum. This new SDSS
measurement is shown in Figure 1. Furthermore, Percival et
al. (2006a) used the BAO scale to determine
$\Omega_m=0.256^{+0.029}_{-0.024}$ (a $\simeq$10\% measurement), which again is
independent of the shape of the power spectrum and thus, independent
of concerns about scale--dependent biasing (assuming it is a smooth function of scale). 

Therefore, the detection of the BAO in the local galaxy distribution
is clear (Figure 1) and recent work demonstrates that it can
deliver robust and competitive measurements of the cosmological
parameters. It is interesting to note that the $\Omega_m$ value
derived from the BAO scale alone is in excellent agreement with
the value of $\Omega_m$ derived from the overall shape of the power spectrum,
i.e., the horizon scale ($\Omega_m h^2$) from the turn--over in the
power spectrum on large scales. Percival et al. (2006b) finds
$\Omega_m=0.22\pm0.04$ from the analysis of the shape of the SDSS DR5 power spectrum,
which includes accurate modeling of the luminosity--dependent biasing
of galaxies.

\section{SDSSII Supernova Survey}

 \begin{figure}[tp]
\plotone{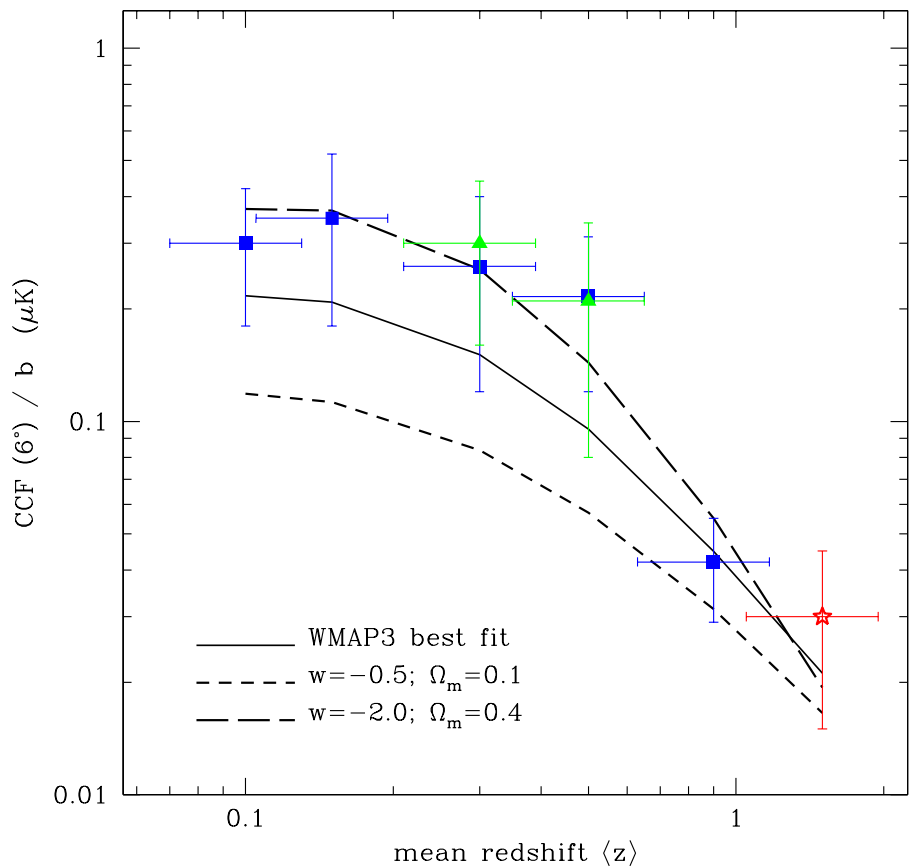}
\caption{Taken from Giannantonio et al. (2006). Summary of ISW detections compared with WMAP3 best fit model (solid line). The blue (squared) points are, in order,  2MASS, APM, SDSS, SDSS high-z, NVSS+HEAO taken from Gaztanaga et al.(2006). The green (triangles) are from Cabre et al. (2006). The red (star) point is the ISW measurement from Giannantonio et al. (2006).  The other lines are $ w = -2 $ (long dashed) and  $ w = -0.5 $ (short dashed) respectively
            \label{ISW}
           }
    \end{figure}

The peak brightness of a supernova Type Ia (SNe Ia) has been calibrated as a standard candle in the Universe to spectacular effect. For example, the first year results from the CFHT Supernova Legacy Survey (SNLS) show that ``$\Omega_M = 0.263 \pm 0.042{\rm (stat)} \pm 0.032{\rm(sys)}$ for a flat $\Lambda$CDM model; and $w = -1.023\pm 0.090{\rm (stat)} \pm 0.054{\rm (sys)}$ for a flat cosmology with constant equation of state $w$ when combined with the constraint from the recent Sloan Digital Sky Survey measurement of baryon acoustic oscillations" (Astier et al. 2006). As detailed in the DETF report, such supernovae are ``the most powerful and best proven technique for studying dark energy. The power of the experiment as reflected in the DETF figure of merit is much better known than for other techniques with the outcome depending on the uncertainties in supernova evolution and in the astronomical flux calibrationÓ. Therefore, we must fully understand the systematic uncertainties associated with using IaÕs as cosmological probes to fully exploit their proven potential. This requires large, homogeneous catalogues of SNe, to ensure Poisson noise or experimental issues (photometric calibrations, etc.) do not limit studies for supernova evolution. 

One such survey is the on--going Sloan Digital Sky Survey II (SDSSII) Supernova Survey. For three months each year (Sept, Oct, Nov), between 2005 and 2007, the SDSSII is dedicated to nightly repeat scans (weather permitting) of  Stripe82, which contains deep ($r\sim24$) multi-color SDSS photometry over 300deg$^2$. Using dedicated mountain-top analysis, the SDSSII imaging data is reduced daily and compared against a deep co-added template of Stripe82 to detect transient objects simultaneously in the SDSS $g,r,i$ passbands. In the first year of operations (Sept to Nov 2005), the SDSS discovered over $2\times10^5$ transients, all of which required visual inspection by 15 scientists worldwide to reject data glitches, asteroids and variable stars. In 2006, a more automated procedure was employed which greatly reduced the number of visually inspected objects by an order of magnitude.The data was also compared against known catalogues of variable stars and AGNs. In 2005, the SDSS found $\simeq12,000$ SNe candidates that were then classified via multicolour light-curve (LC) fitting using theoretical and observational models of different SNe types (II, Ia, Ic, Ib, peculiar, etc.). The LC fitting is $>90$\% successful in correctly identifying Ia's when there are greater than two epochs on the LC, thus greatly improving our efficiency in the spectroscopic follow-up. 
 
In just 90 days in late 2005, the SDSS team discovered 129 Ia's through the spectroscopic follow-up of a subset of SDSS SNe candidates (see Figure \ref{SNe}). At the time of writing, the 2006 campaign has nearly finished and the SDSSII has detected over $300$ Ia's spanning the redshift range $0.05<z<0.4$ as shown in Figure \ref{SNe}. With one more year to go, the SDSSII is on target to spectroscopically--confirm $\simeq500$ low--to--intermediate redshift SNe Ia's. In addition to these spectroscopically--confirmed Ia's, a further $\sim60$ Ia's could be added to the sample based on their known redshifts (from the host galaxy) and the fits to the multicolor LCs, i.e., they all have a high probability of being a Ia. 
 
The spectroscopic follow--up of SDSSII SNe has been achieved using an international network of telescopes. In the US,  the Keck, HET, MDM, Mayall and ARC 3.5m telescopes have been used, while elsewhere, the NTT, NOT, INT, SALT, WHT and Subaru telescope have also made significant contributions. For example, the ESO NTT allocated 17 nights of telescope time for the follow--up of SDSSII SNe in 2006.

Monte Carlo simulations of the final SDSSII SNe survey show it should measure $\Omega_M=0.28\pm 0.02$ and $w=-0.95\pm0.08$, assuming 500 SNe (with the same z--range as Figure \ref{SNe}) and a flat underlying cosmology with $\Omega_M=0.3$ (based on the BAO measurements of the SDSS) and a systematic uncertainty of 0.02mags in the distance modulus to mimic errors in the photometric system. 
However, the greatest gain in the accuracy of these cosmological parameters will be achieved when the SDSSII is combined with the higher--redshift SNLS sample, i.e., the error on $w$ should drop to $\sim6$\% because the SDSSII provides a low redshift anchor for the SNLS Hubble diagram. 

In addition to constraining cosmological parameters, the SDSSII sample will be important for studying the properties of SNe (multi-color LCs, absorption lines) and how they may correlate with redshift (SNe evolution) and with the properties of the host galaxy (progenitor bias, dust extinction). For example, Hachinger et al (2006) published a detailed study of the velocities and equivalent widths of absorption lines in 28 nearby SNe Ia's and discovered that there appears to be 3 subclasses based on the time-averaged rate of decrease of the expansion velocity of the SiII absorption line, i.e., the high-velocity gradient (HVG) class, the low velocity gradient (LVG) class and a FAINT class. It is unclear how these subclasses relate to the recent Sullivan et al. (2006) study of the SNLS SNe, which also find evidence for two subclasses, i.e., ``prompt" Ia's, which are found in young, star--forming galaxies, and ``delayed" Ia's seen in passive (elliptical) hosts. A detailed study of these subclasses and how they relate to the SNe Hubble diagram is critical for the next generation of supernova searches.

\section{Integrated Sachs--Wolfe (ISW) Effect}
 
Another way of probing dark energy is through the late--time Integrated Sachs--Wolfe (ISW) effect, which produces anisotropies in the Cosmic Microwave Background (CMB).  While most of the CMB anisotropies were generated near the last scattering surface (at the moment of recombination), additional anisotropies can be created later via gravitational interactions, e.g., the gravitational redshift (or blueshift) of photons traversing time--evolving potential wells.  During the matter--dominated era, gravitational potentials remain constant and so there is no ISW effect.  However, if the Universe becomes dominated by curvature or dark energy, then additional CMB anisotropies can be created. Unfortunately, it is hard to separate these induced anisotropies (because of the ISW effect) from intrinsic anisotropies, but one way to extract this signal is to correlate the entire CMB anisotropy map with a tracer of the dark matter distribution (Crittenden \& Turok 1996; Afshordi 2004; Peiris \& Spergel 2000) as the primary CMB anisotropies will not be correlated with the late--time matter overdensities.  Using this technique, the ISW effect has now been detected, beyond any doubt, by many authors using a number of different tracers of the underlying matter distribution (Boughn \& Crittenden 2003; Nolta et al. 2003; Afshordi et al. 2004; Fosalba \& Gaztanaga 2004; Scranton et al. 2003; Fosalba et al. 2004;  Padmanabhan et al. 2005; Cabre et al. 2006). 

Regrettably, the statistical power of the ISW effect is limited by cosmic variance and noise from spurious correlations with the intrinsic CMB anisotropies. However, the ISW effect is still important because it provides a direct probe of the effect of dark energy on the growth of cosmic structures in the Universe and can be measured to high redshift ($z>1$), thus testing the dark energy paradigm in a unique and complementary way. For example,  Giannantonio et al. (2006) recently detected (at the 2--2.5 sigma level) the ISW effect at $z\simeq1.5$ through the cross--correlation of high--redshift SDSS quasars and the WMAP3 CMB maps. The detected signal is independent of frequency, as expected for the ISW effect, and robust against the details of the masking and stellar contamination. Without dark energy, they would not expect this detection and their measurement represents the earliest evidence yet for dark energy.

In Figure \ref{ISW}, we compare the Giannantonio et al. (2006) measurement to other ISW detections at lower redshift. As can be seen, all the data are consistent (within the error bars) with the WMAP3 best--fit $\Lambda$CDM cosmology (i.e. $w=-1$), with a best--fit to the high--redshift Giannantonio et al. (2006) data giving $-1.18<w<-0.76$ (assuming $H_0=72\pm8{\rm km\,s^{-1}\,Mpc^{-1}}$). Although, this constraint is not as competitive as other measurements of $w$ (SNe above), remember it is a measurement of dark energy at the redshift of the tracers (i.e. $z>1$), and therefore, offers an ``insurance policy" against missing rapid changes in $w(z)$. Pogosian et al. (2005)  demonstrated this as ``the cross-correlation of Planck CMB data and LSST galaxy catalogs will provide competitive constraints on $w(z)$, compared to a SNAP-like SNe project, for models of dark energy with a rapidly changing equation of state".

\acknowledgements 
Many thanks for my colleagues in the SDSS collaborations for allowing me to show their research and present it as my own both here and at the conference. Thanks to Tom Shanks and Nigel Metcalfe for the patience waiting for this proceeding and their invitation to participate. Finally, thanks to all the Durham locals for making this a very enjoyable conference. 

Funding for the SDSS and SDSS-II has been provided by the Alfred P. Sloan Foundation, the Participating Institutions, the National Science Foundation, the U.S. Department of Energy, the National Aeronautics and Space Administration, the Japanese Monbukagakusho, the Max Planck Society, and the Higher Education Funding Council for England. The SDSS Web Site is http://www.sdss.org/. The SDSS is managed by the Astrophysical Research Consortium for the Participating Institutions. The Participating Institutions are the American Museum of Natural History, Astrophysical Institute Potsdam, University of Basel, University of Cambridge, Case Western Reserve University, University of Chicago, Drexel University, Fermilab, the Institute for Advanced Study, the Japan Participation Group, Johns Hopkins University, the Joint Institute for Nuclear Astrophysics, the Kavli Institute for Particle Astrophysics and Cosmology, the Korean Scientist Group, the Chinese Academy of Sciences (LAMOST), Los Alamos National Laboratory, the Max-Planck-Institute for Astronomy (MPIA), the Max-Planck-Institute for Astrophysics (MPA), New Mexico State University, Ohio State University, University of Pittsburgh, University of Portsmouth, Princeton University, the United States Naval Observatory, and the University of Washington.

\end{document}